\documentclass[11pt]{article}
\usepackage{authblk}
\usepackage{hyperref}
\usepackage[margin=2.5cm]{geometry}
\def\be{\begin{equation}}
\def\ee{\end{equation}}
\def\ba{\begin{eqnarray}}
\def\ea{\end{eqnarray}}

\usepackage{amsfonts}
\usepackage{mathtools}
\usepackage{amsmath}
\usepackage[dvipsnames]{xcolor}
\usepackage{verbatim}
\usepackage{hyperref}
\usepackage{cite}
\usepackage{lineno}
\usepackage{setspace}

\begin{document}

\title{The quantum Hall effect in the absence of disorder}

\author[*]{Kyung-Su Kim}

\author[$\dagger$]{Steven A. Kivelson}
\affil{Department of Physics, Stanford University, Stanford, CA 93405} 
\affil[*]{kyungsu@stanford.edu}
\affil[$\dagger$]{kivelson@stanford.edu}

\date{}
\maketitle

\begin{abstract}
    It is widely held that  disorder is essential to the existence  of a finite interval of magnetic field in which the Hall conductance is  quantized, i.e. for the existence of `plateaus' in the quantum Hall effect. 
    Here, we show that the existence of a quasi-particle Wigner crystal results in the persistence of plateaus of finite extent even in the limit of vanishing disorder.
    Several experimentally detectable features that characterize the behavior in the zero disorder limit are also explored. \\
\end{abstract}

\textbf{Introduction.}--- In the absence of disorder, the electrostatic force exerted by an electric field $\mathbf E$ on an electron fluid 
balances the Lorentz force from a magnetic field $\mathbf B$ so long as the fluid velocity $\mathbf v= c \mathbf E \times\mathbf B/B^2$. 
Consequently, one expects the Hall conductance of a two dimensional electron gas to vary linearly with electron density, $n$,  
i.e. to exhibit the `classical' Hall effect: $\sigma_{xy}= nec/B = (e^2/h) \nu$,   where  the `filling factor' $\nu \equiv n\phi_0/B$ and $\phi_0=hc/e$ is the quantum of flux. 
In contrast,  where the quantum Hall effect arises, the Hall conductance remains constant, $\sigma_{xy}= (e^2/h) \nu^\star$, for a finite range of $\nu$ on either side of $\nu=\nu^\star$, where $\nu^\star$ is 
a rational number corresponding to a 
 quantized filling factor. 
 Combining these observations, it is generally asserted that disorder is necessary for the quantum Hall effect.
 Instead,  we show that the 
  occurrence of a quasi-particle Wigner Crystal (QPWC) \cite{Chen2003microwave, lewis2004nu3WC, Moon2015nu1WC, Hatke2017halfWC,Zhu2010nuthirdQPWC} for sufficiently small $|\nu-\nu^\star|$ results in  a quantized Hall plateau, even in the absence of disorder.  
  To make the considerations concrete, we derive the quantized Hall conductance both in a system with disorder, taking the limit as the disorder strength tends to zero and as the system size to infinity, and in a finite size system with no disorder but with a finite pinning potential at the edges. 
  We also discuss the finite $T$ phase diagram and how  thermal melting of the QPWC restores the classical behavior.

\textbf{Classical Hall effect.}--- The  derivation of the `classical' Hall effect  typically invokes Galilean invariance \cite{girvin1999quantum, girvin2019modern}. 
Although higher-order corrections to effective mass theory do not have Galilean symmetry \cite{landau1988mechanics}, it is straightforward to show that translation invariance alone is sufficient for the Hall conductivity to take its classical value (See Appendix).
The observation of quantized Hall plateau, on the other hand, is commonly associated with the explicit breaking of translational symmetry by disorders \cite{laughlin1981IQHE,tong2016lectures,girvin&Prange, huckestein1995scaling}.


\begin{figure}[t]
\begin{center}
	\includegraphics[scale=0.7]{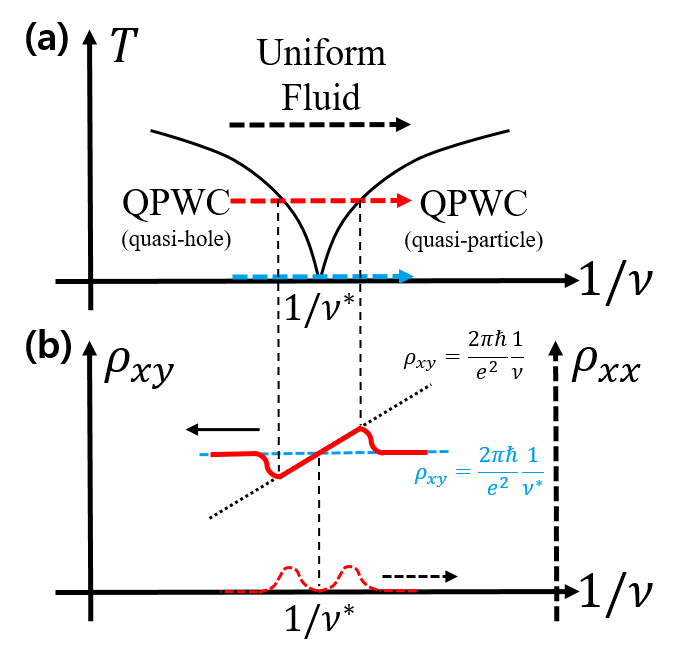}
\caption{\textbf{Finite $T$ phase diagram of `ideal' quantum Hall effect} \\
a) Schematic phase diagram in the neighborhood of $\nu=\nu^\star$ (quantized filling factor) in the limit of vanishing disorder.  b) Resistivity tensor as a function of $\nu$ at three fixed temperatures, as indicated in panel a. At zero temperature (blue) the Hall resistivity is strictly quantized, whereas at high temperature (black) the QPWC has melted and the Hall resistivity takes its classical value. At intermediate temperatures (red), the Hall resistivity shows sharp crossovers from the classical value to the quantized one as $\nu$ moves away from $\nu^*$. Moreover, the longitudinal resistivity $\rho_{xx}$ is very close to zero deep in the QPWC phase or near $\nu^*$ but becomes finite away from those regions.
}
\label{Fig1}
\end{center}
\end{figure}

\textbf{The occurrence of quasi-particle Wigner crystal.}--- In the absence of disorder, however, 
there is one other possible source of a 
quantum Hall plateau: spontaneous breaking of translational symmetry. 
Indeed, the presence of long-range Coulomb interactions makes this inevitable close to a quantized filling factor.  Specifically, assuming the existence of 
a quantum Hall fluid at $\nu=\nu^\star$, for $\delta \nu \equiv (\nu - \nu^\star)$ small, the system consists of a small density 
\be
n_{\textrm{qp}} =  \left( \frac B {\phi_0}\right) \left|\frac e {e^\star}\right| |\delta \nu|
\ee
of quasi-particles or quasi-holes (depending on the sign of $\delta \nu$), where 
$e^*$ is the charge of 
a quasi-particle (i.e. $\left| e^\star/e \right| = 1$ for the integer quantum Hall effect and a suitable rational fraction for the fractional quantum Hall effect).  
Thus, for $|\delta \nu|$ sufficiently small, the spacing between quasi-particles is large compared to the magnetic length, $\ell = \sqrt{\phi_0/2\pi B}$, and they behave as classical particles with fixed guiding-centers.
The result is that the quasi-particles form a Wigner crystal below a transition temperature,
\be
\label{eq:T_QPWC}
T_{\textrm{qpwc}} = A_{\textrm{wc}}\ \frac{(e^{*})^2}{4 \pi\epsilon} \left[ \pi n_{qp} \right]^{1/2}\sim |\delta \nu|^{1/2}\ ,
\ee
where $\epsilon$ is the dielectric constant and $A_{\textrm{wc}} \approx 1/128$ \cite{GrimesAdams, MorfWCmelting, ChuiWCmeltiing}. 
For $T<T_{\textrm{qpwc}}$ the quasi-particle Wigner crystal (QPWC) has quasi-long-range order and a finite shear modulus;  it only has true long-range order at $T=0$. The resulting phase diagram for the `ideal' (i.e. zero disorder) quantum Hall system in the neighborhood of $\nu^\star$ is thus as shown in Fig. 1(a).  
Importantly, at low $T$ in the crystalline state, the lowest lying current carrying excitations are interstitials, 
whose density is exponentially activated as
\be
n_{\textrm{int}} \sim \exp[-\Delta/T] \ \  {\rm with } \ \ \Delta = A_{\textrm{int}}\ \frac{ (e^\star)^2}{4 \pi \epsilon}\left[ n_{\textrm{qp}} \right]^{1/2},
\ee
where $A_{\textrm{int}} \approx 0.136$ \cite{FHM1979defects, cockayne:defectEnergy}.

The width of the QPWC phase can be estimated using the composite fermion approach\cite{Jain1989CF}.  
In the absence of Landau level mixing, it was estimated in Ref.\cite{Lam&Girvin} that an electron WC has lower energy than any quantum Hall liquid for $\nu < \nu_0$ where $1/7 < \nu_0 < 1/5$;  Landau level mixing will generally tend to lead to a larger value of $\nu_0$.  In the mean-field treatment of composite fermions with  $2m$ flux quanta attached to each electron, an electron state with  $\nu= (p +f)/[1 +2m(p+f)]$ corresponds to a state with $p$ fully filled and one  $f$-filled Landau level of composite fermions. 
($m$ and $p$ are integers and $0 < f< 1$.)  
This reasoning leads to the conclusion that if there is a stable FQH liquid at $\nu^\star = p/[1+2mp]$, there should be a stable QPWC for
\be
\frac {p-\nu_0}{1+2m(p-\nu_0)} < \nu <\frac {p+\nu_0}{1+2m(p+\nu_0)} .
\label{range}
\ee
For instance, taking $\nu_0\approx 1/7$ \cite{Lam&Girvin} the QPWC associated with the $\nu=1/3$ plateau would extend over the range $1/3 - 0.018< \nu < 1/3 + 0.014$ in the absence of disorder.

\begin{figure}[t]
\begin{center}
	\includegraphics[scale=0.5]{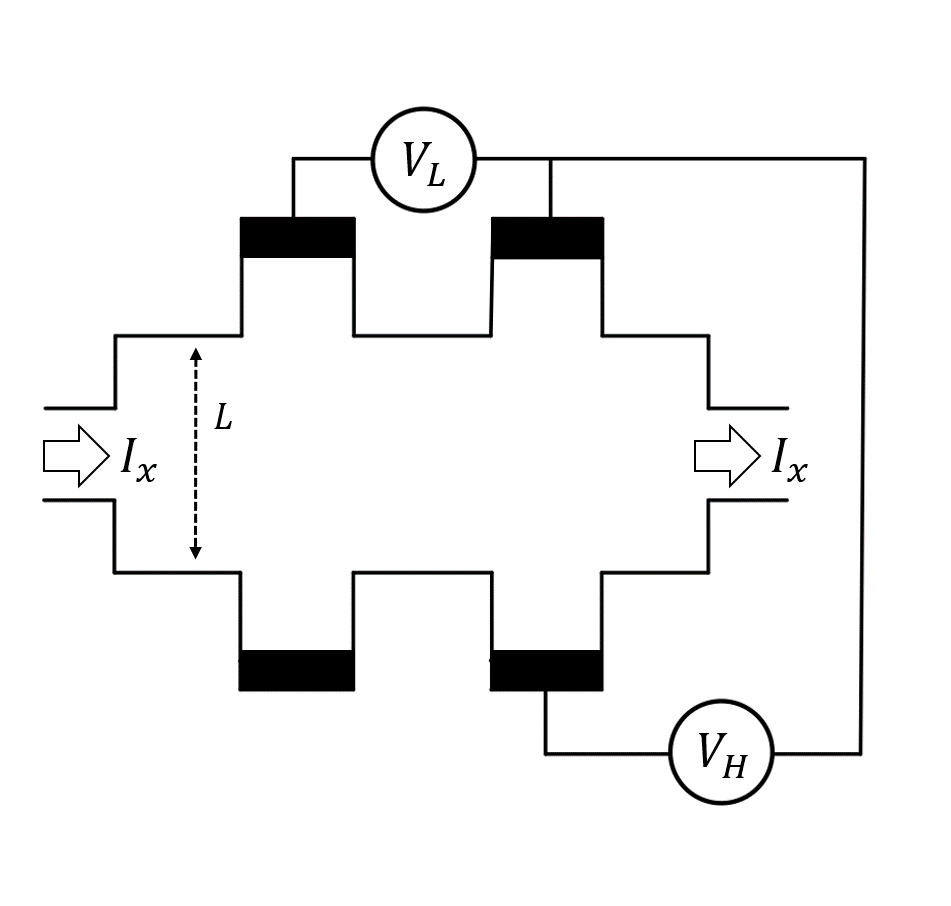}
\caption{\textbf{Schematic of a Hall bar} \\ As the current flows through the current leads, the longitudinal and Hall voltages are measured at the voltage leads. $L$ is the linear size of the sample.}
\label{Fig2}
\end{center}
\end{figure}

\textbf{The pinning of QPWC and quantum Hall effect.}--- Because a Wigner crystal of electrons is insulating in the pinning regime \cite{Giamarchi1998PRL,Giamarchi2001WC, Huse2000WC, reichhardt2016WCdepinning, monarkha2012WC}, the DC conductivity of the QPWC must come exclusively from the quantum Hall condensate. 
As in the case of any charge-density wave (CDW) \cite{FL1, FL2, LR, gruner1988CDW, fisher1985sliding, sneddon1982sliding}, 
  the one subtlety concerns the possible contribution of a sliding state of the QPWC.  We analyze this problem in two distinct ways:

Firstly, we consider a sequence of `real' systems with 
 disorder strength, $D$, as $D\to 0$.
As is well known, disorder in two dimension both destroys long-range CDW order \cite{imry&Ma1975}-- i.e. it rounds the transition at $T_{\textrm{qpwc}}$ -- and pins the CDW \cite{FL1, FL2}.  
This means that for a QPWC with $D\neq 0$, there is a quantized Hall response in a finite range of filling factors: 
 $\sigma_{xy} \to (e^2/h)\ \nu^\star$ as $T\to 0$ for $\nu \in [\nu^*-\delta\nu_-,\nu^*+\delta\nu_+]$ where $\delta\nu_\pm(D)$ approach finite positive values deduced from Eq. \ref{range} as $D\to 0$.
On the other hand, 
for field strengths greater than 
a critical field, $E^\star$, 
the QPWC depins (slides) and contributes to the current.
The depinning field in the weak disorder limit can be estimated using the arguments of Lee and Rice \cite{FL1, LR}:
\be
E^\star(D) \approx 
 \frac {\epsilon}{(e^\star)^{3}}\ D^2, \
\ee
where $D^2$ is the mean square pinning energy per 
unit cell of the QPWC.  
In other words, 
for ever smaller $D$, 
in order to remain in the linear response regime, we must restrict  measurements of the conductivity to an increasingly smaller range of field strengths, $V_H/L=E \ll E^\star(D)$, where $V_H$ is the 
Hall voltage and $L$ is the width of the device as in Fig. \ref{Fig2}. 
We note that even though the critical current density $j_x^*=\rho_{xy}E^*$ goes to $0$ as $D\rightarrow 0$, the critical current $I_x^* \sim LD^2$ 
{could} remain finite if we simultaneously take $L\to \infty$ in an appropriate manner.
To obtain a sense of the expected magnitude of the critical current, we make a crude estimate  $D^2 \sim \left(\frac{ee^*}{4\pi \epsilon \ell }\right)^2\times \left({n_\textrm{i} \ell^2}\right)$ where $n_\textrm{i}$ is the density of impurities. 
In the cleanest two dimensional electron system as of today \cite{chung2020recordquality} (in a $d=30$nm width GaAs quantum well) $n_\textrm{i}= 3\times 10^7\textrm{cm}^{-2}$;
for a 2D electron density $n=1\times 10^{11}\textrm{cm}^{-2}$ and $B=12$T, which corresponds to $\nu=1/3,$ 
this estimate yields a critical Hall voltage $V_H^* \sim 34 \textrm{mV}$ and correspondingly, for a sample with linear size $L_x=0.4 \textrm{cm}$, to a critical current $I_x \sim 0.43\mu \textrm{A}$.

Alternatively, consider measuring the Hall response in an `ideal' ($D=0$) sample of finite size as in Fig. 2.  
Here the current enters and leaves through current leads at either end, and the longitudinal and Hall voltages are measured at the voltage leads on the sides of the device. 
Because the QPWC has a finite shear modulus, it cannot smoothly flow through the current leads at either end  of the device -- it is pinned by the edges of the device.
On the other hand, if the current exceeds a critical value, the crystal will be sheared, and a portion of the QPWC will begin to flow through the device.  
As in the disordered case, this means that the QPWC is pinned for small enough $E$.
Now, assuming  finite pinning potential at the edges, the critical field 
scales to 0 as $E^*\sim 1/L$ in the thermodynamic limit, but the critical Hall voltage and current remain finite even in the limit $L\rightarrow \infty$: $V_H^*=\rho_{xy}I_x^*=E^*L$. 
Again, so long as measurements are carried out with $I_x \ll I_x^*$, a quantized Hall response will be observed (
to an exponential accuracy with small corrections, $\delta \sigma_{xy} \sim \exp[-L/\ell]$). 
Importantly, the width of the plateau, $[\delta\nu_-+\delta\nu_+]$, approaches a constant as $L\to\infty$.

\textbf{The melting of QPWC.}--- The role of the QPWC can (in principle) be unambiguously inferred from the thermal evolution even in the presence of weak disorder.  
For weak enough disorder, one expects a sharp crossover in conductivity 
at $T=T_{\textrm{qpwc}}(\delta\nu)$: 
 for $T> T_{\textrm{qpwc}}$, the Hall response should be close to its `classical' 
 value, while for $T < T_{\textrm{qpwc}}$ the quantum Hall effect should be seen up to exponentially small corrections \cite{Jiang1991Activation}  proportional to the concentration of interstitials,  $n_{\textrm{int}}$.
 Since $T_{\textrm{qpwc}}(\delta\nu)$ vanishes as $\delta\nu \to 0$ as given in Eq. \ref{eq:T_QPWC}, 
 the finite $T$ Hall response should exhibit a peculiar reentrant behavior as a function of $\nu$ as shown in Fig. 1b.
The temperature below which this reentrant behavior can be seen is evaluated to be $T_{\textrm{qpwc}}\approx 0.02 \textrm{K}$, where we take $\nu=1/3-0.009\ (=1/3-\delta\nu_{-}/2)$ and $n=10^{11}\textrm{cm}^{-2}$ in Eq. \ref{eq:T_QPWC}. 

\section*{Appendix}
Here we prove that the Hall conductivity takes the classical value $\sigma_{xy}=nec/B=(e^2/h)\nu$ as discussed in the main article. Consider a gauge-invariant many-body Hamiltonian 
\be
H=\sum_iK\left[ \mathbf p_i+\frac ec\mathbf A(\mathbf r_i,t)\right] + V\left[\left\{ \ \mathbf r_j \right\} \right]
\ee 
in which $V$ is invariant under (continuous) translation, $\mathbf r_i \to \mathbf r_i + \mathbf a$, where $\mathbf r_i$ is the (2d) position of the $i$-th electron and $\mathbf p_i$ the conjugate momentum.  We consider an infinite cylinder with periodic boundary conditions in the $\hat y$ direction,
a uniform perpendicular magnetic field, $\mathbf B=B\hat z$, and an electric field $\mathbf E =E \hat y$, which 
can be represented, in an appropriate gauge, as  $\mathbf A(\mathbf r,t) =\left [-cEt +Bx\right]\hat y$. 
The electric field dependent term in $H$ can be removed by a suitable transformation to a co-moving frame with velocity $\mathbf v=(E/B) \hat x$. 
Specifically, let $\mathbf{r}_j '=\mathbf r_j -c(E/B)t \hat x$ and correspondingly $\mathbf p_i'= \mathbf p_i$, then $i \hbar \frac{\partial}{\partial t}=i \hbar \frac{\partial}{\partial t'}+c\frac{E}{B}\sum_i p_{i,x}'$ and the Hamiltonian transforms to:
\be
H'=\sum_jK\left[ \mathbf p_j'+\frac ec Bx_j' \hat y  \right] +V\left[\left\{ \mathbf r_j' \right\} \right]-c\frac{E}{B}\sum_j p_{j,x}',
\ee
which is time-independent. 
The current density in the $x$-direction is simply proportional to the center of mass velocity, $j_x' = -n e \langle \dot X\rangle$
(where $n$ is the electron density), which 
must likewise be time-independent.  
The total momentum in $y$-direction,  $P_y'=P_y$, is conserved, while for fixed $P_y'$, 
$\langle K'\rangle $ would diverge with time as $|X'| 
\to \infty$ 
for any $\left< \dot X' \right> \neq 0$.
Consequently, $\left < \dot X' \right> =0$ and  $j_x'=0$.
However, the current density transforms as $\mathbf j'(\mathbf r')=\mathbf j(\mathbf r)+e c\frac EB n(\mathbf r)\hat x$, and hence, $j_x =-necE/B$, which gives classical Hall conductivity.

 \section*{Acknowledgement}
 We thank S. Hartnoll and C. Murthy for constructive criticism.  This work was supported, in part, by the U. S. Department of Energy (DOE) Office of Basic Energy Science, Division of Materials Science and Engineering at Stanford under contract No. DE-AC02-76SF00515.
 
 \section*{Contributions}
Both authors contributed essentially to all aspects of this paper.

\section*{Competing Interests}
The authors declare no competing financial or non-financial interests.

\end{document}